\documentclass[%
reprint,
onecolumn,
 amsmath,amssymb,
 aps,
]{revtex4-2}

\usepackage{graphicx}
\usepackage[hidelinks]{hyperref}

\usepackage{dcolumn}
\usepackage{bm}
\usepackage{amsmath}
\usepackage{makecell}
\usepackage{ulem}

\begin{document}


\title{Complex network approach to the turbulent velocity gradient dynamics: high- and low-probability Lagrangian paths}

\author{C.~J.~Keylock${}^{1,2}$}
\email{C.J.Keylock@lboro.ac.uk}
\author{M.~Carbone${}^{2}$}

\affiliation{${}^{1}$Loughborough University, School of Architecture, Building and Civil Engineering, Loughborough, Leicestershire, U.K.\\
${}^{2}$Theoretical Physics I, University of Bayreuth, Germany}

\date{\today}

\begin{abstract}
Understanding the dynamics of the turbulent velocity gradient tensor (VGT) is essential to gain insights into the Navier-Stokes equations and improve small-scale turbulence modeling. However, characterizing the VGT dynamics conditional on all its relevant invariants in a continuous fashion is extremely difficult. Hence, in this paper, we represent the VGT Lagrangian dynamics using a network where each node represents a unique flow state. The sign and relative magnitude ranking of the invariants resulting from a Schur decomposition of the VGT determines the discrete flow state. This approach enables us to discern the role of VGT non-normality and its production rates and to study how the VGT transitions from one state to another in a simplified fashion. Our analysis reveals intriguing features of the resulting network, such as the clustering of the commonly visited nodes where the eigenvalues of the VGT are real, in the proximity of the Vieillefosse tail. We then relate our complex network approach to the well-established VGT discretization based on the sign of its principal invariants, $Q$ and $R$, and its discriminant, $\Delta$. To this end, we separate the shortest paths on the network (geodesics) based on the $Q$-$R$ region to which their starting and arrival nodes belong. The distribution of the length of intra-region geodesics, with starting and arrival nodes belonging to the same $Q$-$R$ region, exhibits a distinct bimodality in two regions of the $Q$-$R$ plane, those in which the deviatoric part of the pressure Hessian introduces complexity to the VGT dynamics. Such bimodality is associated with 
infrequently visited nodes having to follow a long, low probability path to drastically change the state of the VGT compared to other flow states that can acquire the necessary characteristics without changing their sign for $Q$ or $R$. 
The geodesics analysis also reveals how the typically clockwise path on the $Q$-$R$ diagram arises. We complement the geodesics approach by examining random walks on the network, showing how the VGT non-normality and the associated production terms distinguish the shortest commuting paths between different $Q$-$R$ regions.
\end{abstract}

\maketitle

\section{Introduction}
Understanding the properties of turbulent fluid flows at small scales is crucial for gaining insights into the mechanisms by which flows transfer and dissipate energy \citep{Vassilicos2015,Carbone2020,Johnson2020} and how this relates to extreme events \citep{Buaria2019,VelaMartin2022}.
This is important both for modelling the velocity gradients \citep{meneveau11} and for understanding the physical mechanisms that lead to constraints on the smoothness of the underpinning Navier-Stokes equations \citep{frisch95,ohkitani16,Drivas2023,Miller2023}
\begin{align}
\bm{\nabla \cdot u} = 0, &&
\frac{\partial \bm{u}}{\partial t} + (\bm{u \cdot \nabla})\bm{u} = -\bm{\nabla} p + \nu \nabla^2 \bm{u}.
\label{eq.NS}
\end{align}
Here $\bm{u}$ is the three-dimensional velocity vector, $t$ is time, $p$ is the pressure field divided by the constant density and $\nu$ is the kinematic viscosity.
The smallest scales and extreme events are conveniently addressed from the viewpoint of the velocity gradient tensor (VGT), $\bm{A}\equiv\nabla\bm{u}$.
The governing equation for the VGT follows by taking the spatial gradient of the Navier-Stokes equations \eqref{eq.NS}
\begin{align}
\mbox{tr}(\bm{A}) = 0, && 
\frac{\partial \bm{A}}{\partial t} + (\bm{u \cdot  \nabla})\bm{A} = -\left(\bm{A}^2-\frac{\mbox{tr}(\bm{A}^2)}{3}\bm{I}\right) - \bm{H} + \nu \nabla^2 \bm{A},
\label{eq.NSgrad}
\end{align}
where $\bm{H}$ is the deviatoric part of the pressure Hessian, $\bm{H}\equiv \bm{\nabla\nabla} p -\nabla^2p/3\bm{I}$, and $\bm{I}$ is the identity matrix in three dimensions.
Equation \eqref{eq.NSgrad} has been studied extensively, especially concerning its reduced-order modeling \citep{vieillefosse84,tsinober97,Cantwell1993,ooi99,chacin00}, as reviewed by \citep{meneveau11} and \citep{wilczek24}.
Reduced-order modeling is crucial when equation \eqref{eq.NSgrad} is interpreted from a Lagrangian perspective, that is when one considers the dynamic evolution of the VGT along a fluid particle trajectory. Both the pressure Hessian and viscous terms are nonlocal functionals of the velocity gradient \citep{ohkitani95}, and they are not uniquely determined by the knowledge of $\bm{A}$ at a single point in space. From this single-particle/single-time viewpoint, equation \eqref{eq.NSgrad} is unclosed, and the unclosed/nonlocal terms $\bm{H}$ and $\nu\nabla^2\bm{A}$ require modeling.

Of particular importance in this field is the restricted Euler model \citep{Vieillefosse1982,cantwell92}, consisting of equation \eqref{eq.NSgrad} without the deviatoric pressure Hessian and viscous contributions. The restricted Euler model provides qualitative insights into the actual turbulent VGT dynamics and also helps introduce a few quantities crucial for the VGT dynamics, extensively studied in the literature, such as the VGT principal invariants, the dissipation rate, enstrophy, etc.
The restricted Euler model yields two coupled ordinary differential equations for the evolution of the second, $Q = -\mbox{tr}(\bm{A}^{2})/2$, and third, $R = -\mbox{det}(\bm{A})$, principal invariants of the VGT, where $\mbox{tr}(\ldots)$ is the trace and $\mbox{det}(\ldots)$ is the determinant:
\begin{align}
\frac{d Q}{d t} = -3 {R}, &&
\frac{d R}{d t} = \frac{2}{3} {Q}^{2}.
\label{eq.QR}
\end{align}
Solutions to the model \eqref{eq.QR} follow lines of constant discriminant,
\begin{align}
\Delta = Q^{3} + 27 R^{2}/4,    
\end{align}
and thus, proceed from negative $R$ to positive $R$ and negative $Q$, with a timescale $\tau = |\Delta|^{-1/6}$. This results in a finite-time blow-up for almost all initial conditions along the Vieillefosse tail \citep{Vieillefosse1982,vieillefosse84}.
Contrarily, this finite-time blow-up is not observed in the numerical solution of the Navier-Stokes equations, in which the phase-space trajectories explore the $Q$-$R$ diagram \citep{Chen1990,Girimaji90}. Still, the statistical distribution of the phase-space trajectories display signatures of the divergence along the Vieillefosse tail, as hinted by the restricted Euler model, being skewed towards positive $R$ and negative $Q$ (see figure \ref{fig.QRsketch}). For a Taylor Reynolds number of $\mbox{Re}_{\lambda} = 71$, it has been shown that a typical orbit around the $Q$-$R$ diagram takes approximately thirty Kolmogorov times \citep{martin98}, which for that simulation equated to about three eddy turnover times.

Remarkably, within the restricted Euler model, the evolution of the VGT principal invariants $Q$ and $R$ is independent of the other VGT invariants \citep{martin98}, as in equation \eqref{eq.QR}. Extending the restricted Euler model to the full VGT requires consideration of the deviatoric part of the pressure Hessian $\bm{H}$ and the viscous effects stemming from $\nu\nabla^2\bm{A}$ in \eqref{eq.NSgrad}, and how their contributions, conditional on the local VGT state, make the VGT invariants interact among each other.
Several studies have proposed phenomenological models and statistical closures for these nonlocal terms \citep[e.g.][]{Girimaji90,chevillard08,wilczek14,johnson16} which can qualitatively capture the VGT statistics, and recent machine learning methods have permitted accurate predictions and modeling \citep{Buaria2023,Carbone2024}. These modeling approaches require relatively advanced computational tools and very large datasets. In this work, we propose a complementary description of the VGT, based on projecting the VGT Lagrangian dynamics on a complex network. This approach allows us to gain an intuitive and lower-complexity understanding of small-scale turbulence based on discrete VGT states, crucial for improving e.g., flow classifications \citep{Sharma2021}. Furthermore, we can employ the tools from network theory \citep{Latora2017} to characterize the VGT dynamics indirectly, by studying the properties of the associated complex network.

\section{Complex network approach to the VGT dynamics}

Modeling the VGT Lagrangian dynamics remains challenging even in the case of homogeneous, isotropic turbulence (HIT). This is due to the high dimensionality of the problem since the phase space consists of eight continuous variables (the VGT independent components) with their intricate interplay and time correlations. To circumvent some of these difficulties, in this paper, we propose a different approach based on projecting the dynamics of the VGT onto a discrete network. We define the network nodes by partitioning the phase space into discrete regions according to the sign and relative magnitudes of relevant VGT invariants.
From a statistical viewpoint, classic Lagrangian VGT modeling aims at reproducing the relevant joint distribution functions of the VGT, while the present complex network approach amounts to binning those joint distributions on a coarsely partitioned phase space. This considerably reduces the degrees of freedom and makes it easier to explore the interplay of multiple flow features.

Using complex networks to study nonlinear phenomena is an important strand of nonlinear physics \citep[e.g.][]{newman03,xiao21}. In the context of turbulence research, network-based approaches are not numerous and have been used to study, for example, velocity induction by mutual vortex interaction in 2D turbulence \citep{nair15,taira16}
coherent structures \citep{Scarsoglio2016} and time irreversibility \citep{Iacobello2023}. Those studies defined the weighted adjacency matrix in terms of physical-space realisations of the flow. Conversely, we define the nodes based on the VGT phase-space variables. In particular, we employ scalar invariants that can be formed from the VGT \citep{Pope1975,Itskov2015,Leppin2020,Carbone2023} and VGT components obtained from its complex Schur decomposition \citep{k18}.

\subsection{Discrete variables}
The VGT scalar invariants provide insight into the VGT state since they relate to enstrophy, dissipation, production terms, etc. They are even more relevant in statistically isotropic turbulence in which the VGT single-time statistics can be parameterized in terms of only five independent invariants \citep{Itskov2015,Carbone2023}.
We can make a variety of choices on the independent invariants, we begin with the ones complementing the $Q$-$R$ picture \eqref{eq.QR}. 
A classical decomposition consists of splitting the VGT into its symmetric part, the strain rate $\bm{S}=(\bm{A}+\bm{A}^*)/2$, and its anti-symmetric part, the rotation rate $\bm{\Omega}=(\bm{A}-\bm{A}^*)/2$, where the asterisk indicates complex conjugation and matrix transposition.
Using this decomposition one can rewrite the principal invariants as
\begin{align}
Q &= \frac{1}{2}\left(\Vert\bm{\Omega}\Vert^{2} - \Vert\bm{S}\Vert^{2}\right),\label{eq.Q}\\
R &= -\mbox{det}(\bm{S}) - \mbox{tr}(\bm{\Omega}^{2}\bm{S}),
\label{eq.R}
\end{align}
where $\Vert\ldots\Vert$ denotes the Frobenius norm. The second invariant is thus related to the enstrophy and dissipation rate while the third relates to the strain production and enstrophy production.
For a more detailed characterization of the VGT state, we adopt a complex Schur decomposition \citep{k18}, namely
\begin{align}
    \bm{A} = \bm{U}\bm{T}\bm{U}^*, &&
    \bm{T} = \bm{L}+\bm{N},
\end{align}
where $\bm{U}$ is a unitary matrix, $\bm{L}$ is a diagonal matrix of eigenvalues and $\bm{N}$ is the upper-triangular non-normality matrix \citep{henrici62}.
Then, by defining the matrices $\bm{B}$ and $\bm{C}$ and decomposing them into their Hermitian and anti-Hermitian parts
\begin{align}
    \bm{B} = \bm{U}\bm{L}\bm{U}^{*} = \bm{S}_B + \bm{\Omega}_B, &&
    \bm{C} = \bm{U}\bm{N}\bm{U}^{*} = \bm{S}_C + \bm{\Omega}_C,
    \label{eq.Hermitedecomp}
\end{align}
we write the second-order invariants in terms of the normal enstrophy, normal strain and non-normality
\begin{align}
\label{eq.Om}
\Vert\bm{\Omega}\Vert^{2} &= \Vert\bm{\Omega}_{B}\Vert^{2} + \Vert\bm{\Omega}_{C}\Vert^{2}\\
\Vert\bm{S}\Vert^{2} &= \Vert\bm{S}_{B}\Vert^{2} + \Vert\bm{\Omega}_{C}\Vert^{2}.
\label{eq.strain}
\end{align}
Here $\bm{S}_B = (\bm{B}+\bm{B}^*)/2$ denotes the Hermitian part of $\bm{B}$ while $\bm{\Omega}_B = (\bm{B}-\bm{B}^*)/2$ denotes its anti-Hermitian part (with the same decomposition for $\bm{C}$).
Using \eqref{eq.Hermitedecomp} we can rewrite the strain production and enstrophy production as
\begin{align}
\label{eq.detS}
-\mbox{det}(\bm{S}) &= -\mbox{det}(\bm{S}_{B}) -\mbox{det}(\bm{S}_{C}) +\mbox{tr}(\bm{\Omega}_{C}^{2}\bm{S}_{B}) \\
\mbox{tr}(\bm{\Omega}^{2}\bm{S}) &= \mbox{tr}(\bm{\Omega}_{B}^{2}\bm{S}_{B})-\mbox{det}(\bm{S}_{C}) +\mbox{tr}(\bm{\Omega}_{C}^{2}\bm{S}_{B})
\label{eq.trOmS}
\end{align}
where the three terms on the right-hand side of \eqref{eq.detS} represent the normal strain production, $-\mbox{det}(\bm{S}_{B})$, non-normal production, $-\mbox{det}(\bm{S}_{C})$, and interaction production, $\mbox{tr}(\bm{\Omega}_{C}^{2}\bm{S}_{B})$, while the additional term on the right-hand side of \eqref{eq.trOmS} is the normal enstrophy production, $\mbox{tr}(\bm{\Omega}_{B}^{2}\bm{S}_{B})$.
Through the decomposition \eqref{eq.Hermitedecomp} we can deal with physically relevant quantities defined from the VGT, clearly distinguishing between normal terms (those related to $\bm{B}$) and non-normal terms (those related to $\bm{C}$). 

We use the invariants and associated quantities described above to define the discrete state variable $\overrightarrow{v}(t)$ which labels the signs of the invariants and their relative magnitude as
\begin{align}
\nonumber
\overrightarrow{v}(t) &=\\
\nonumber
&\mbox{sign}\left[Q,R,\Delta,-\mbox{det}(\bm{S}),
\mbox{tr}(\bm{\Omega}^{2}\bm{S}), 
-\mbox{det}(\bm{S}_{C}),
\mbox{tr}(\bm{\Omega}_{C}^{2}\bm{S}_{B})
\right]\\
\nonumber
&\mbox{rank}\left[
\Vert\bm{\Omega}_{B}\Vert^{2},
\Vert\bm{\Omega}_{C}\Vert^{2},
\Vert\bm{S}_{B}\Vert^{2}
\right]\\
&\mbox{rank}\left[-\mbox{det}(\bm{S}_{B}),
-\mbox{det}(\bm{S}_{C}),
\mbox{tr}(\bm{\Omega}_{C}^{2}\bm{S}_{B}),
\mbox{tr}(\bm{\Omega}_{B}^{2}\bm{S}_{B})
\right],
\label{eq.state}
\end{align}
where $\mbox{sign}[\dots]$ indicates all the possible sign combinations of the arguments and  $\mbox{rank}[\dots]$ indicates descending ordering of the arguments. Notice that the signs of $-\mbox{det}(\bm{S}_{B})$ and $\mbox{tr}(\bm{\Omega}_{B}^{2}\bm{S}_{B})$ do not need to be included in the definition \eqref{eq.state} since they are equal to, and opposite to, the sign of $R$, respectively.

Each node in the complex network is formed based on its own unique combination of signs/rankings of the terms in \eqref{eq.state}. For example, the $Q$-$R$ decomposition, classically used to characterize the VGT and that we adopt for much of the paper, may be defined in terms of the signs of $Q$, $R$ and $\Delta$, yielding the six regions sketched in Fig.~\ref{fig.QRsketch}. 
The total number of possible states \eqref{eq.state} is reduced by physical or mathematical constraints. For example, when $\Delta < 0$ then $\Vert\bm{\Omega}_{B}\Vert^{2} = 0$ and is therefore ranked the smallest of the terms given in the second row of \eqref{eq.state}. It then follows that $\mbox{tr}(\bm{\Omega}_{B}^{2}\bm{S}_{B}) = 0$ and is ranked the smallest of the production terms. Moreover, if $R > 0$, then $-\mbox{det}(\bm{S}_{B}) > 0$ and, thus, from equations \eqref{eq.detS} and \eqref{eq.trOmS} if the non-normal production and interaction production are positive then so must be $-\mbox{det}(\bm{S})$. In addition, if $|\mbox{det}(\bm{S}_{C})| > |\mbox{tr}(\bm{\Omega}_{C}^{2}\bm{S}_{B})|$ then $-\mbox{det}(\bm{S})$ will still be positive even if $\mbox{tr}(\bm{\Omega}_{C}^{2}\bm{S}_{B}) < 0$, etc. Taking into account all these constraints gives $N = 848$ possible states for the VGT based on the definition of our state vector in \eqref{eq.state}. 
These form the nodes for the full network and the region exit``X'' network defined below although, in practice, nodes that were visited extremely infrequently, or had no edges between nodes in different regions, respectively, were excluded from the network as detailed below.

\subsection{Constructing the complex network}
A network (graph) $\mathcal{G}$, without self-loops, comprises a set of $N$ nodes (vertices), $\mathcal{V}$, edges $\mathcal{E}$, and weights, $\mathcal{W}$.
The possible configurations of the discrete state $\overrightarrow{v}$ define the nodes, while the weights follow from the probability of transitioning from a node to any of the other nodes. We determine that transition probability based on Lagrangian tracking within a flow simulation. We consider time realizations of the VGT $\bm{A}(t)$, thus of the discretized state vector $\overrightarrow{v}(t)$, along an ensemble of fluid particle trajectories. Then, we count the transitions from node $i$ at time $t_i$, $\overrightarrow{v}(t_i) = \overrightarrow{v}_{i}$, to node $j\ne i$ at time
$t_j>t_i$, $\overrightarrow{v}(t_j) = \overrightarrow{v}_{j}$, with $t_j-t_i$ a multiple of the sampling step of the Lagrangian trajectories, $\Delta t$.
With this Lagrangian approach the transition probability acts as a lens for the dynamics: only accepting edges between vertices if the transition probability exceeds a threshold simplifies the dynamics as the threshold increases. We adopted the maximum resolution that the data permitted to provide as much detail as possible.

The weights may be considered simultaneously via the $N \times N$ adjacency matrix $\bm{M}$ \citep{Latora2017}
\begin{equation}
m_{ij} = \left \{ \begin{matrix} w_{ij} & \textrm{if }\, (i,j)\, \in \, \mathcal{E}\\
0 & \textrm{otherwise}
\end{matrix}  \right.
\label{eq.mdef}
\end{equation}
where the $w_{ij}$ are the observed probabilities of the the transition from node $i$ to node $j$, with $\sum_{i}\sum_{j} m_{ij} = 1$.
In the absence of symmetry for these weights the network is \textit{directed} and the out-degree, $D_{O}(i) \equiv \sum_{j} m_{ij}$ will not equate to the in-degree, $D_{I}(j) \equiv \sum_{i} m_{ij}$. To capture this asymmetry certain metrics can be applied separately to both the out-degree and in-degree. For example, the out-degree eigenvalue centrality, $C_{\lambda,O}$, is based on the right eigenvector for the largest eigenvalue of $\bm{M}$ while the in-degree variant, $C_{\lambda,I}$, uses the equivalent left eigenvector.

\begin{figure}
\includegraphics[width=.6\textwidth]{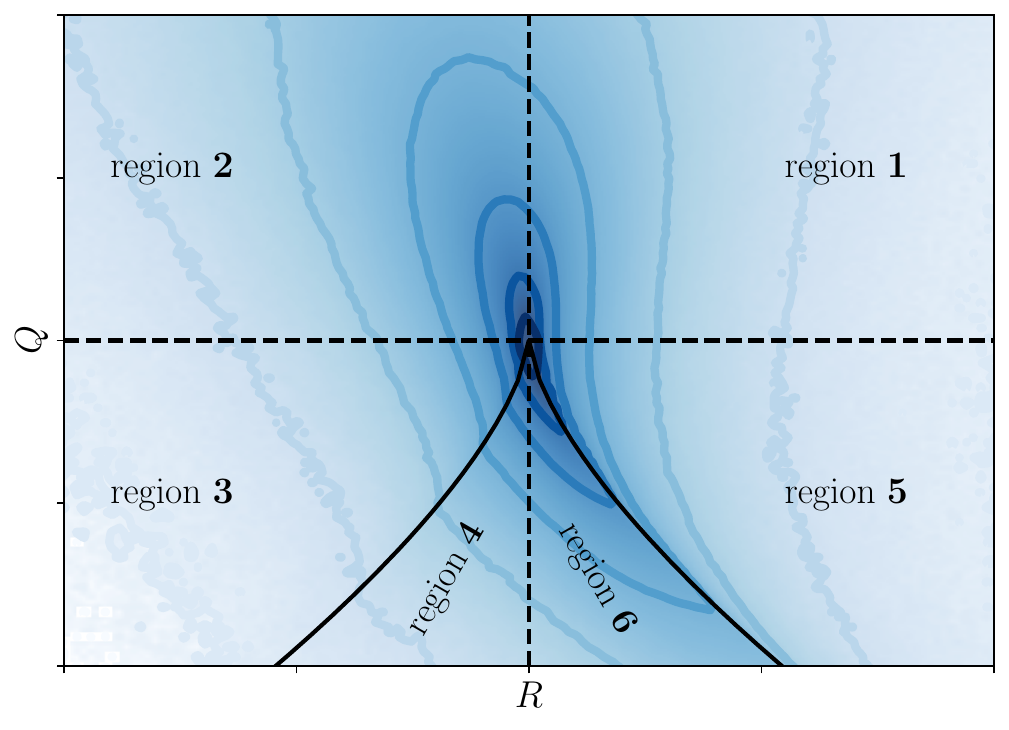}
\caption{Contours of the joint distribution function of the VGT principal invariants, $Q$ and $R$, with the six regions identified based on the signs of $Q$, $R$ (dashed lines) and $\Delta$ (solid line) indicating the Vieillefosse tails.}
\label{fig.QRsketch}
\end{figure}

Given the importance of the principal invariants for studying the VGT dynamics \citep{betchov56,Vieillefosse1982,cantwell92,nomura98,Carbone2022}, we also focus explicitly on the transition probabilities between regions of the $Q$-$R$ plane that are defined based on the signs of $Q$, $R$ and $\Delta$ as shown in Fig.~\ref{fig.QRsketch}. Thus, generalizing our notation from \eqref{eq.mdef}, we consider the adjacency matrix weights as $m_{i,\alpha;j,\beta}$, where the $(i,j)$ subscripts label each node and $(\alpha,\beta) \in \{1,\ldots,6\}$ indicate that node $i$ belongs to region $\alpha$ and node $j$ belongs to region $\beta$. To focus on the flux around the $Q$-$R$ diagram and to exclude the dynamics within a region, we then have the alternate adjacency matrix, $\bm{M}^{X}$, with weights
\begin{equation}
m^{X}_{i,\alpha;j,\beta} = \left \{ \begin{matrix} m_{i,j} & \textrm{if } \, \alpha \ne \beta\\
0 & \textrm{otherwise}
\end{matrix}  \right.,
\label{eq.mXdef}
\end{equation}
where the ``X'' superscript highlights that this is the adjacency matrix for edges that exit and, thus, connect regions of the $Q-R$ diagram.

\section{Method}
We employ a simulation of HIT at a Taylor Reynolds number of $\mbox{Re}_{\lambda} = 433$ in a $1024^3$ computational domain with Lagrangian tracking of fluid particles from the Johns Hopkins Turbulence Database \citep{yili,Yu2012_JHTDB}. Within the computational domain, we chose 27 initial seeding locations along each axis and we track the $27^{3}$ fluid particles for 240 Kolmogorov times, $\tau_{\eta}$ with a temporal sampling step of $\Delta t = \tau_{\eta}/20$. In constructing the network, when the flow packet transitions from the state indexed by $i$, to that indexed by $j$, we increment the value of $w_{ij}$ by one. After tracking all particles for the entire time series duration we normalize the $w_{ij}$ by $\sum_{i,j} w_{ij}$ such that the adjacency matrix, $\bm{M}$ contains normalized weights summing to unity. 

Notice that the flow state transition probabilities we analyse and those that follow from the Lagrangian dynamics of the VGT differ since we  exclude self-loops as is typically done in network analysis. Hence, our focus is explicitly on the transitions from one flow state to another; the residence time in a flow state $i$ before the transition to a distinct node $j$ is not incorporated into our approach. This is similar to symbolic sequence analyses of turbulence time series \citep{k14}. Thus, if we had four flow states 
and measured the flow state at six regularly sampled points in time to obtain $\{1,2,2,3,4,4\}$ on one occasion and $\{1,1,1,2,3,4\}$ on another, both would give the same directed network with the same edge weights.

\subsection{Network metrics}
Classic network analysis commonly draws upon the notion of \textit{centrality} to evaluate the importance of a particular node. 
We employ the eigenvector and betweenness centralities to characterize the complex network and then focus in greater detail on the properties of the shortest paths (geodesics) and random walks on the network. 
To capture the extent to which important nodes are clustered with other important nodes we make use of the out-degree eigenvector centrality, $C_{eO}$ \citep{bon72}
and a metric of asymmetry between the out-degree and in-degree, $C_{eI}$, centralities:
\begin{equation}
\kappa_{e}(i) = \frac{C_{eO}(i)-C_{eI}(i)}{C_{eO}(i)+C_{eI}(i)},
\label{eq.kappa}
\end{equation}
where $C_{eO}$ employs the right eigenvectors of $\bm{M}$ and $C_{eI}$ the left eigenvectors, and as these are equivalent for an undirected network, one would obtain $\kappa_{e} = 0$ in that case. The Perron-Frobenius theorem ensures that the largest eigenvalue of $\bm{M}$, $\lambda_{1,1}$, is positive and, thus, all the elements of the first column vector of the eigenvector matrix, which defines $C_{eO}$ (or $C_{eI}$), are also positive. We adopt the standard normalization of the eigenvectors such that $\sum_{i=1}^{N} C_{eO}(i) = 1$. The inclusion of an ``X'' superscript with $C_{eO}$, $\kappa_{e}$ etc., indicates that the underpinning adjacency matrix is $\bm{M}^{X}$ rather than $\bm{M}$.

Further insight is provided by the geodesics on the network. Given the shortest path (geodesic) between vertices $i$ and $j$, the betweenness centrality for vertex $k$ is the fraction of shortest paths between the $i$ and $j$ that pass through $k$ \citep{freeman77,borgatti06}.
With $\phi(i,j)$ the total number of shortest paths and $\phi(i,j|k)$ those passing through $k$, the betweenness centrality is
\begin{equation}
C_{b}(k) = \frac{\phi(i,j|k)}{(N-1)(N-2)\phi(i,j)},
\end{equation}
where $\phi(i,j|k) = 0$ if $k \in \{i,j\}$. Hence, $C_{b}(k)$ is the extent to which node $k$ controls, on average, the pair-wise interaction between other nodes. To consider the structure of the network in greater detail, we focus explicitly on the shortest paths and describe them both in terms of the nodes and the regions of the $Q-R$ diagram they pass through.
A geodesic starting from node $i$ in region $\alpha$, ending at node $j$ in region $\beta$ and passing through a node $k$ in region $\gamma\ne\alpha$ can be written as the set of the vertices in the order they are visited
\begin{equation}
   \mathcal{S}(i,\alpha; j,\beta) = \{v_{i,\alpha}, v_{k_1,\gamma_{1}}, \ldots, v_{k_{\xi},\gamma_{\xi}},\ldots, v_{k_T,\gamma_{T}},v_{j,\beta}\},
   \label{eq.geodesic}
\end{equation}
where vertices $v_{k_1}$ to $v_{k_T}$ are the intermediary nodes that are visited, and these are in intermediary regions, $\gamma_{\xi}$, which may equate to $\alpha$ or $\beta$. Hence, the length of the shortest path may be expressed in terms of the number of edges traversed, $\ell_{S}(i,j) = T+1$.
With definition \eqref{eq.geodesic} an intra-region geodesic has $\alpha = \beta$ and an inter-region geodesic has $\alpha \ne \beta$. In the majority of cases, therefore, the $\gamma_{\xi}$ for an intra-region geodesic are all equal to $\alpha$ (and $\beta$), while an inter-region geodesic between non-adjacent $Q$-$R$ regions (see Fig.~\ref{fig.QRsketch}) will necessarily have at least one $\gamma_{\xi} \ne \{\alpha, \beta\}$.

Other measures of nodal separation for directed networks include the hitting and commute distances \citep{boley11}. The former is the average distance between nodes $i$ and $j$ based on a random walk on the network, while the latter is the average distance from node $i$ to $j$ and then back to $i$. Hence, the hitting and commute distances move away from a focus on geodesics to consider all possible pathways on the network. To parallel our approach to the length of geodesics, the length of a commuting random walk is expressed as the average total number of nodes visited, $\overline{\ell}_{C}(i,j) = \overline{T} + 1$, with the bar denoting the average over a large number of random walks (while a geodesic is a single path realization on the network by definition).

We have now outlined some of the underpinning physics of the VGT dynamics, defined our discrete network of flow states and described the primary methods from the complex network literature we will adopt. We can now analyse how the VGT dynamics impact upon the associated complex network structure, and how the network metrics highlight aspects of the VGT dynamics. 

\section{Results}
As noted above, in total, all combinations of signs and relative magnitudes of the terms in the state vector \eqref{eq.state} gave $N = 848$ nodes. However, four of those nodes were visited so infrequently that their probability of occurrence could not be precisely determined. The largest in-degree value for these four 
nodes, when translating the very small probability into a frequency, corresponded to a single occurrence (from 4800 samples over 240 $\tau_{\eta}$) on 0.5\% of trajectories. This was five times less frequent than the fifth least visited node. Hence, these four vertices were subsequently excluded from the network to give $N = 844$ nodes that were used in analysis.

\subsection{Degree distribution}
\begin{figure*}
\includegraphics{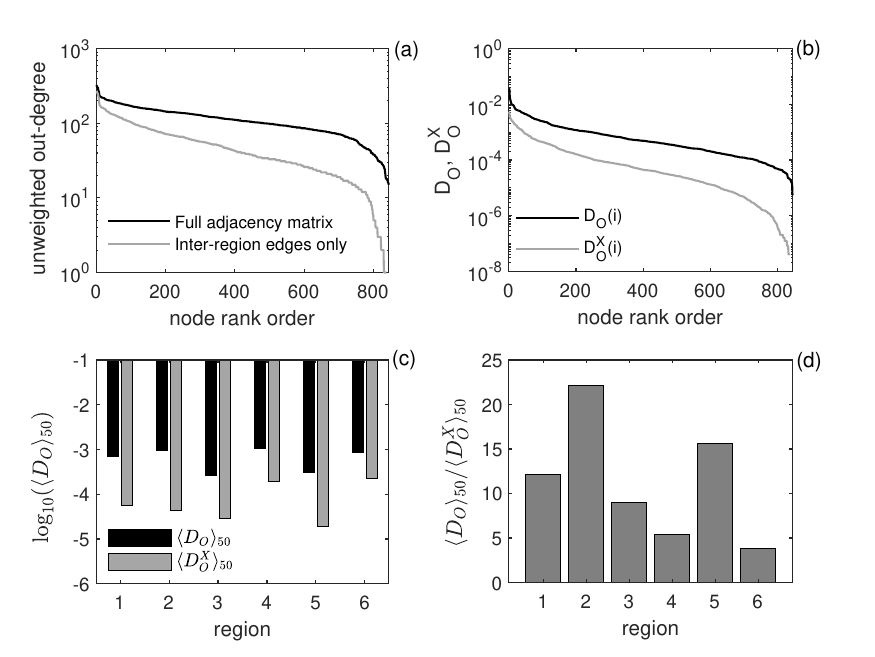}
\caption{The out-degree distributions, $D_{O}$ and $D_{O}^{X}$, for the unweighted network in which positive entries of $\bm{M}$ are replaced by ones, (a) and the full weighted network (b). The median values of $D_{O}$ and $D_{O}^{X}$ for each $Q$-$R$ region are reported on a log-scale in (c) and the ratio of these values is in (d). Here and throughout $\langle\ldots\rangle_{50}$ indicates the median, and the “X” superscript indicates that the analysis is based on $\bm{M}^X$ with edges connecting the $Q$-$R$ regions.}
\label{fig.OutD}
\end{figure*}

The degree distribution of a network provides a signature of the network's structure with, for example, the preferential attachment model leading to a power-law degree distribution and a small-world network \citep{BA99}. Such analyses are typically undertaken for an unweighted network such that the degree distribution reflects the network topology rather than both the topology and the strength of the edges. Hence, in panel (a) of Fig.~\ref{fig.OutD} we first show the unweighted degree distribution for both the full adjacency matrix (the unweighted equivalent of $\bm{M}$, in which each positive value is replaced by a $1$) and for the adjacency matrix for the inter-region edges (the unweighted equivalent of $\bm{M}^{X}$). This may be compared to the distributions of $D_{O}$ and $D_{O}^{X}$ in panel (b). In all cases, the distribution is exponential from approximately the one-hundredth ranked node to about node 750, before a steep drop in the degree. This is particularly large for $D_{O}^{X}$ and its equivalent in panel (a) due to the number of nodes whose edges are oriented to adjusting the dynamics of the tensor for a constant $Q$, $R$ and $\Delta$. The impact of the weights is to steepen the slope of the exponential part of the distribution, rather than to change the qualitative nature of the overall topology from exponential to, for example, a power-law.

Region-based median values of $D_{O}$ and $D_{O}^{X}$ are given in Fig.~\ref{fig.OutD}(c) on a logarithmic scale, and the ratio between these values is given in (d) to highlight the differences between the two forms of adjacency matrix. Regions 4 and 6 (where $\Delta < 0$), as well as region 2 ($Q > 0$, $R < 0$), have the largest average values of $D_{O}$ but while this pattern persists for $D_{O}^{X}$ for the $\Delta < 0$ regions, the average value for region 2 declines by a factor of 20, as shown in panel (d). Given a typical clockwise motion around the $Q-R$ diagram (e.g. Fig. 6 of \citep{k19}), this large value for region 2 in (d) implies a relatively infrequent transition from region 2 to region 1. That is, where $Q > 0$ and $R < 0$ (hence, the normal enstrophy production is positive), many of the changes in the dynamics are about reconfiguring the properties of the vortex in a manner that does not impact the signs of the normal enstrophy production and normal strain production.
The second largest value in panel (d) is for region 5 also implying significant flow reconfiguration within that region. Given the clockwise nature of the typical Lagrangian path on the $Q$-$R$ plane,  and the number of nodes in region 5 and 6 (Table \ref{table.QR}). The reduction from 239 nodes in region 5 to 60 nodes in region 6 is driven by $\Vert\boldsymbol{\Omega_{B}}\Vert^{2} = 0$ when $\Delta \le 0$. So, the flow needs to reconfigure internally within region 5 from vertices that represent relatively high rankings for the normal enstrophy or its production to ones where these terms are sufficiently small to be connected to otherwise similar nodes in region 6.

\subsection{Centrality metrics}

\begin{figure*}
\includegraphics{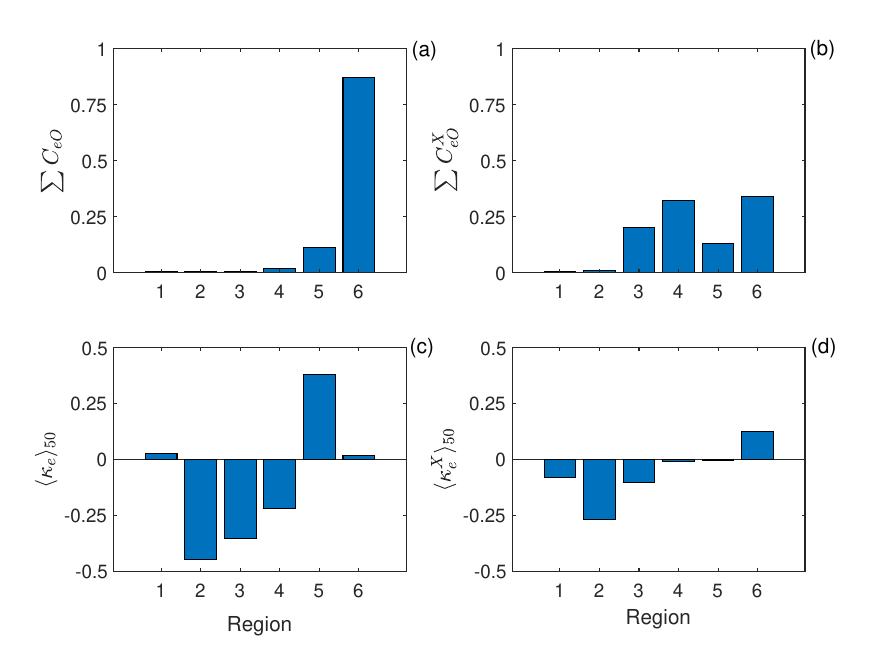}
\caption{Summary values for the centrality metrics, $C_{eO}$ (top row) and $\kappa_{e}$ (bottom row) for adjacency matrices $\bm{M}$ (left column) and $\bm{M}^{X}$ (right column) and for each region of the $Q-R$ diagram (on the horizontal axis).
$\langle\ldots\rangle_{50}$ indicates the median.}
\label{fig.centrality}
\end{figure*}

The strongly connected nature of $\bm{M}$ means that the Perron-Frobenius theorem holds allowing for the existence of $C_{eO}$ and $C_{eI}$. To ensure that this was also the case for $\bm{M}^{X}$ it was necessary to remove seven nodes from region 3 and three from region 5 where either the out-degree, $D_{O}(i) = \sum_{j} m^{X}_{ij}$ or the in-degree, $D_{I}(j) = \sum_{i} m^{X}_{ij}$ was zero (because these vertices are only connected to other nodes in the same region). Summary values for the centrality statistics for both $\bm{M}$ and $\bm{M}^{X}$ are given in Fig.~\ref{fig.centrality} and Table \ref{table.QR}. The results are reported as a function of the region of the $Q$-$R$ diagram, labeled by $\alpha$, as shown in Fig.~\ref{fig.QRsketch}, with $N_{\alpha}$ indicating the number of nodes in region $\alpha$. Figure \ref{fig.centrality}(a) shows that 98\% of $\sum C_{eO}$ arises in regions 5 and 6 (those that form the right Vieillefosse tail \citep{Vieillefosse1982}). Concerning $\sum C_{eO}^{X}$ instead, the results are a strong function of $Q$, with values less than 2\% for regions 1 and 2 where $Q > 0$, between 12\% and 20\% for regions 3 and 5 ($Q < 0, \Delta > 0$) and at least 30\% for regions 4 and 6 ($Q < 0, \Delta < 0$), as displayed in Fig.~\ref{fig.centrality}(b). Thus, the overwhelming tendency is for nodes with high out-degree centrality to cluster in the regions adjacent to the right Vieillefosse tail. However, the difference in the results between  $\sum C_{eO}$ and $\sum C_{eO}^{X}$ indicates that a high proportion of these nodal connections are within-region, and when these are excluded, there is greater similarity in network structure for the pairs of regions with a different sign of $R$ but the same signs for $Q$ and $\Delta$. 

\begin{table}
\begin{center}
\begin{tabular}{clcc}
 Region $\alpha$ & $N_{\alpha}$ & $\langle{C}_{b} \rangle_{50}$ & $\langle{C}_{b}^{X} \rangle_{50}$\\
 & & $(\times 10^{-3})$ & $(\times 10^{-3})$\\
\hline
1 & 123& 6.2 (0.85) & 6.5 (0.72) \\
2 & 123 & 9.0 (0.83) & 6.2 (0.72) \\
3 & 239, \textit{232} & 3.1 (0.86) & 4.7 (0.46) \\
4 & 60& 4.0 (0.95) & 9.4 (0.92) \\
5 & 239, \textit{236} & 2.4 (0.89) & 5.5 (0.46) \\
6 & 60& 7.4 (0.93) & 8.6 (0.90) \\
\end{tabular}
\end{center}
\caption{Median values of the betweenness centralities, $C_{b}$ and $C_{b}^{X}$ for the nodes belonging to each $Q$-$R$ region, $\alpha$ (see Fig.~\ref{fig.QRsketch}). The number of nodes in a region is $N_\alpha$, while the values in italics for $N_{\alpha}$ are the adjusted number of nodes per region for $\bm{M}^{X}$.
The values in brackets for the betweenness centralities' columns indicate the proportion of nodes in the region over which the median is calculated, (i.e. excluding the nodes that do not lie on any shortest path).}
\label{table.QR}
\end{table}

The results for $\kappa_{e}$ and $\kappa_{e}^{X}$, shown in Fig.~\ref{fig.centrality}c-d, also display a marked contrast. While the sign of $\kappa_{e}$ is given by the sign of $R$ (nodes for $R > 0$ have greater out-degree than in-degree eigenvector centrality), $\kappa_{e}^{X}$ is positive only in region 6. Hence, when the focus is on the transitions between regions of the $Q$-$R$ diagram, nodes that are strong donors of transition probability flux only cluster in region 6. In contrast, Fig.~\ref{fig.centrality}d shows that region 2 exhibits a preferential clustering of the nodes that are strong receipt vertices. This reinforces the result in Fig. \ref{fig.OutD}d that region 2 is focused on intra-region flow reconfiguration.

Table \ref{table.QR} shows the values for the median betweenness centralities $\langle C_{b}\rangle_{50}$ and $\langle C_{b}^{X}\rangle_{50}$ accompanied by the proportion of nodes in the region that lie on a geodesic (stated in brackets). The value for $N_{\alpha}$ decreases for regions 3 and 5 when $\bm{M}^{X}$ is analysed as there were seven (respectively, three) nodes for which all the edges were to nodes within the same region. The median values quoted were determined over the nodes lying on geodesics since, in the case of $\langle C_{b}^{X}\rangle_{50}$ for regions 3 and 5, with only 46\% of nodes lying on a shortest path, the median would go to zero. In other words, the low median $C_{b}$ values for region 3 and 5 would be even lower if one determined the median over all nodes in the region. As with the eigenvector centrality, region 6 plays an important role for both networks, although in contrast to the eigenvector centrality results, the $Q > 0$ regions (1 and 2) are much more important, with region 2 having the largest value for $\langle C_{b}\rangle_{50}$. Thus, from its eigevector centrality, region 2 has a relatively low clustering of its dynamically important vertices compared to region 6, but from Table \ref{table.QR} these feature on a large proportion of geodesic pathways. Table \ref{table.QR} also shows that $\langle C_{b}^{X}\rangle_{50}$ is a decreasing function of $N_{\alpha}$. This is the expected pattern since the presence of more vertices in a given region implies fewer geodesics through a given node on average. The fact that $\langle C_{b}\rangle_{50}$ departs from this simple pattern highlights that in region 2 there are intra-node geodesics that contribute significantly to the betweenness centrality without directly impacting the Lagrangian circulation around the $Q-R$ diagram. This result is in agreement with the eigenvector centrality ratios in Fig.~\ref{fig.OutD}d.

\subsection{Intra-region geodesics}

\begin{figure*}
\includegraphics{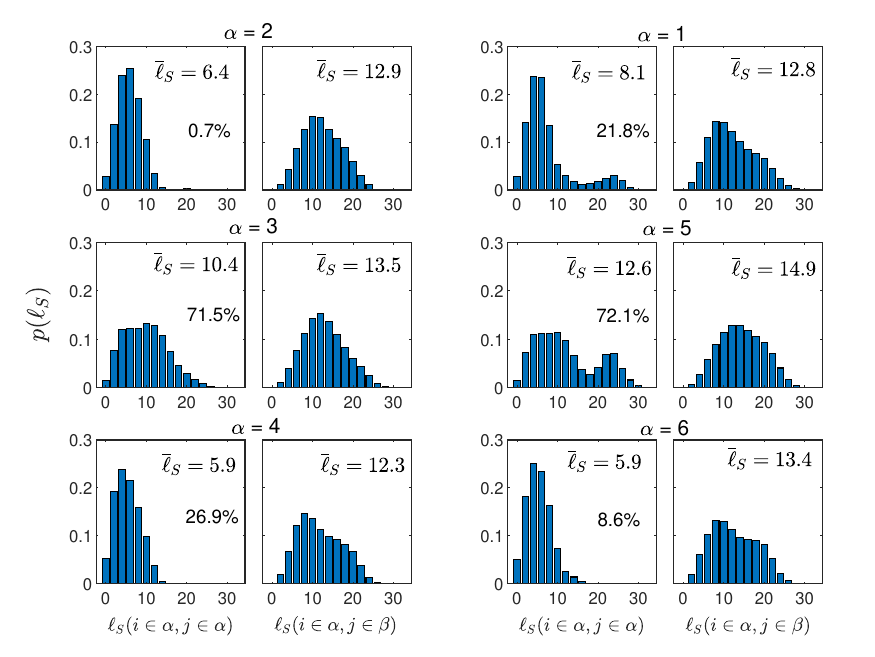}
\caption{Probability of the shortest path length $\ell_{S}(i,j)$ for each $Q$-$R$ region. The intra-region geodesics, reported on the left-hand panel of each region, start from node $i$ belonging to region $\alpha$ and end at node $j$ also belonging to region $\alpha$. The inter-region geodesics, displayed on the right-hand panel of each region, start from node $i$ belonging to region $\alpha$ and end at node $j$ belonging to region $\beta\ne\alpha$. We quote the mean geodesic length $\overline{\ell}_{S}$ in each panel. The percentage of intra-region geodesics that explore other regions, i.e., from region $\alpha$ to region $\gamma\ne\alpha$ and back to $\alpha$, is shown in each left-hand panel. }
\label{fig.ellSN}
\end{figure*}

Looking at the geodesics in greater detail, Fig.~\ref{fig.ellSN} compares the length of the shortest path $\ell_{S}(i,j)$ for different $Q$-$R$ regions, distinguishing between intra- and inter-region geodesics (the left- and right-hand panels for each region). The ordering of the panels by region reflects the structure in Fig.~\ref{fig.QRsketch}. The mean geodesic lengths stated in the left-hand panels of each region in Fig.~\ref{fig.ellSN} indicate that for the intra-region geodesics, the path length is primarily a function of $N_{\alpha}$ with a secondary dependence upon $R$ for $\Delta > 0$, similar to the results for $\langle C_{b}^{X}\rangle_{50}$ in Table \ref{table.QR}. The percentage stated in each of the left-hand panels is the proportion of the intra-region geodesics that include nodes in other regions. That is, a geodesic \eqref{eq.geodesic}
with $\alpha = \beta$ but some of the $\gamma_\xi$ values not equal to $\alpha$. Thus, region 2 is the most self-contained as fewer than 1\% of intra-region geodesics involve vertices outside this region. This then leads to a small mean geodesic length $\overline{\ell}_{S}(i,j) = 6.4$ for $\alpha = 2$, consistent with our interpretation of the structure of region 2 as involving significant flow reconfiguration within the region from the eigenvector and betweenness centralities.

However, the intra-region geodesics for regions $\alpha=4$ and $\alpha=6$ also have low mean lengths despite a significant proportion of them visiting other intermediate regions $\gamma \ne \alpha$. Investigating this, it was found that such geodesics only involve the other region with $\Delta < 0$. Hence, in this particular sense, $\Delta < 0$ is more relevant for the VGT dynamics than the sign of $R$ combined with the sign of $\Delta$, as the high amount of interchange between regions 4 and 6 implies they are almost acting as one. It may also be noted that the $26.9\%$ of intra-region geodesics of region 4 with $\gamma \ne \alpha$ all transition via region 6 and thus have a preferential direction opposite to the average clockwise Lagrangian circulation around the $Q-R$ diagram (although in agreement with the paths for the restricted Euler model). 
The intra-region geodesics starting from regions 2, 4 and 6 have relatively short lengths, as shown by the narrow distribution of $\ell_{S}$ in Fig.~\ref{fig.ellSN}. Conversely, the length distribution of the geodesics starting from region 3 displays a significant right tail, such that the intra-region geodesic lengths are surprisingly close to the inter-region geodesic lengths for this region. For example, 4\% of the intra-region geodesics for region 3 have length $\ell_{S}(i,j) > 20$, and, comparably, 10\% of the region 3 inter-region geodesics have length $\ell_{S}(i,j) > 20$.
More surprisingly, the intra-region geodesics length distributions for regions 1 and 5 (i.e.~$R > 0$, $\Delta > 0$) are marked by a genuine bimodality. In both cases, the left mode exhibits a similar shape to the equivalent region at $R < 0$, while the right mode has a value of $\ell_{S}(i,j) \sim 24$. The large volume of the upper mode for region 5 results in the probability of $\ell_{S}(i,j) > 20$ being greater for the intra-region geodesics (22\%) than the inter-region geodesics (17\%). It was found that this upper mode of long geodesics is driven by cases that undertake a full orbit of the $Q$-$R$ diagram to reach another flow state in the same region. Therefore, in regions 1 and 5 there is a set of vertices characterized by a relatively low probability of transitioning to other parts of the same region directly.

\begin{table}
\begin{center}
\begin{tabular}{ccrrrr}
 \makecell{Sign of \\$-\mbox{det}(\bm{S}_{C})$} & \makecell{Sign of \\ $\mbox{tr}(\bm{\Omega}_{C}^{2}\bm{S}_{B})$} & \makecell{Exit \\ nodes} & \makecell{Entrance \\ nodes} & \makecell{Whole \\ region 5} & \makecell{All \\ nodes}\\
\hline
- & - & 36\% & 56\% & 11\% & 10\%\\
- & + & 34\% & 9\% & 34\% & 33\%\\
+ & - & 15\% & 31\% & 13\% & 14\%\\
+ & + & 15\% & 4\% & 42\% & 43\%\\
\end{tabular}
\end{center}
\caption{Relative frequency of joint states of the non-normal production, $-\mbox{det}(\bm{S}_{C})$, and the interaction production, $\mbox{tr}(\bm{\Omega}_{C}^{2}\bm{S}_{B})$, for exit and entrance nodes on intra-region geodesics that start from region 5 and traverse every region of the $Q$-$R$ diagram.
We also show those relative frequencies computed over all nodes in region 5 and all nodes in the network.}
\label{table.reg5}
\end{table}

Further inspection of the results highlights a difference between regions 1 and 5 concerning the formation of these distinct sets of vertices. 
For region 1, intra-region geodesics with length $\ell_{S}(i,j) > 20$ exhibit a major change in either the non-normality or its production along the path. Thus, they either start from nodes $i$ where  $\Vert\bm{\Omega}_{C}\Vert^{2}$ is ranked third in magnitude of the second-order terms (see the discrete state definition \eqref{eq.state}) and terminate at nodes $j$ where it is ranked first and/or, at $i$, $-\mbox{det}(\bm{S}_{C})$ and $\mbox{tr}(\bm{\Omega}_{C}^{2}\bm{S}_{B})$ were ranked third and fourth, respectively, with $-\mbox{det}(\bm{S}_{B})$ second but at $j$ the latter was now ranked fourth as the production terms involving the non-normal tensor increased in prominence. In all these cases, the normal enstrophy, $\mbox{tr}(\bm{\Omega}_{B}^{2}\bm{S}_{B})$ was ranked first. Hence, increasing $\Vert\bm{\Omega}_{C}\Vert^{2}$ innately  within region 1 is difficult, and an orbit of the $Q$-$R$ diagram is required to increase tensor non-normality significantly.

Although it is both regions 1 and 5 that exhibit intra-region bimodality, the mechanism explaining this is different in region 5, where $Q < 0$ and, therefore the ordering of the normal enstrophy and normal straining swaps compared to region 1. It was found that bimodality was less about the change in relative magnitude of $\Vert\bm{\Omega}_{C}\Vert^{2}$ and more about the changes in signs of the production terms. This is shown in Table \ref{table.reg5}, which looks at the signs of the non-normal production and the interaction production for the nodes that are the last in the region for a long intra-region geodesic as it exits region 5 at the start of the geodesic, compared to those where it re-enters towards the end of the geodesic. The results show that the nodes through which the geodesics re-enter region 5 preferentially feature a negative interaction production, $\mbox{tr}(\bm{\Omega}_{C}^{2}\bm{S}_{B}) < 0$ (87\% of cases compared to 24\% in general for nodes in region 5 and the network as a whole). This leads to a net movement along the geodesic towards both $\mbox{tr}(\bm{\Omega}_{C}^{2}\bm{S}_{B}) < 0$ and $-\mbox{det}(\bm{S}_{C}) < 0$ (36\% for exit nodes and 56\% as the flow re-enters region 5 compared to 11\% for region 5 in general and 10\% for all nodes in the network). 

Although the exact mechanism explaining the long intra-region paths associated with $R > 0$ and $\Delta > 0$ regions differs between regions 1 and 5, it is noteworthy that the bimodality arises here and that its explanation involves either the magnitude of the non-normality or the signs of the non-normal and interaction production. This is because it is these two regions where the deviatoric/anisotropic pressure Hessian contribution to (\ref{eq.NSgrad}) results in complex dynamics. This can be seen in conditional trajectory analysis \citep{ooi99} that separates contributions from the restricted Euler terms (\ref{eq.QR}), viscous terms and the deviatoric pressure Hessian (an example of this analysis is shown in Figure 3 of \citep{zhou15}). Given that the restricted Euler terms may be written in terms of the VGT eigenvalues \citep{meneveau11} that, by definition form the normal part of the tensor, the non-normal contributions to the tensor are associated preferentially with $\bm{H}$. Thus, in order to be coupled to a frequently traversed pathway out of region 1 or 5, an appropriate dynamical signature due to $\bm{H}$ is required. If a flow state does not have this, it cannot acquire it without this inducing a change in $Q$, $R$ or $\Delta$ and, thus a change in region.

\subsection{Inter-region geodesics}

\begin{figure*}
\includegraphics{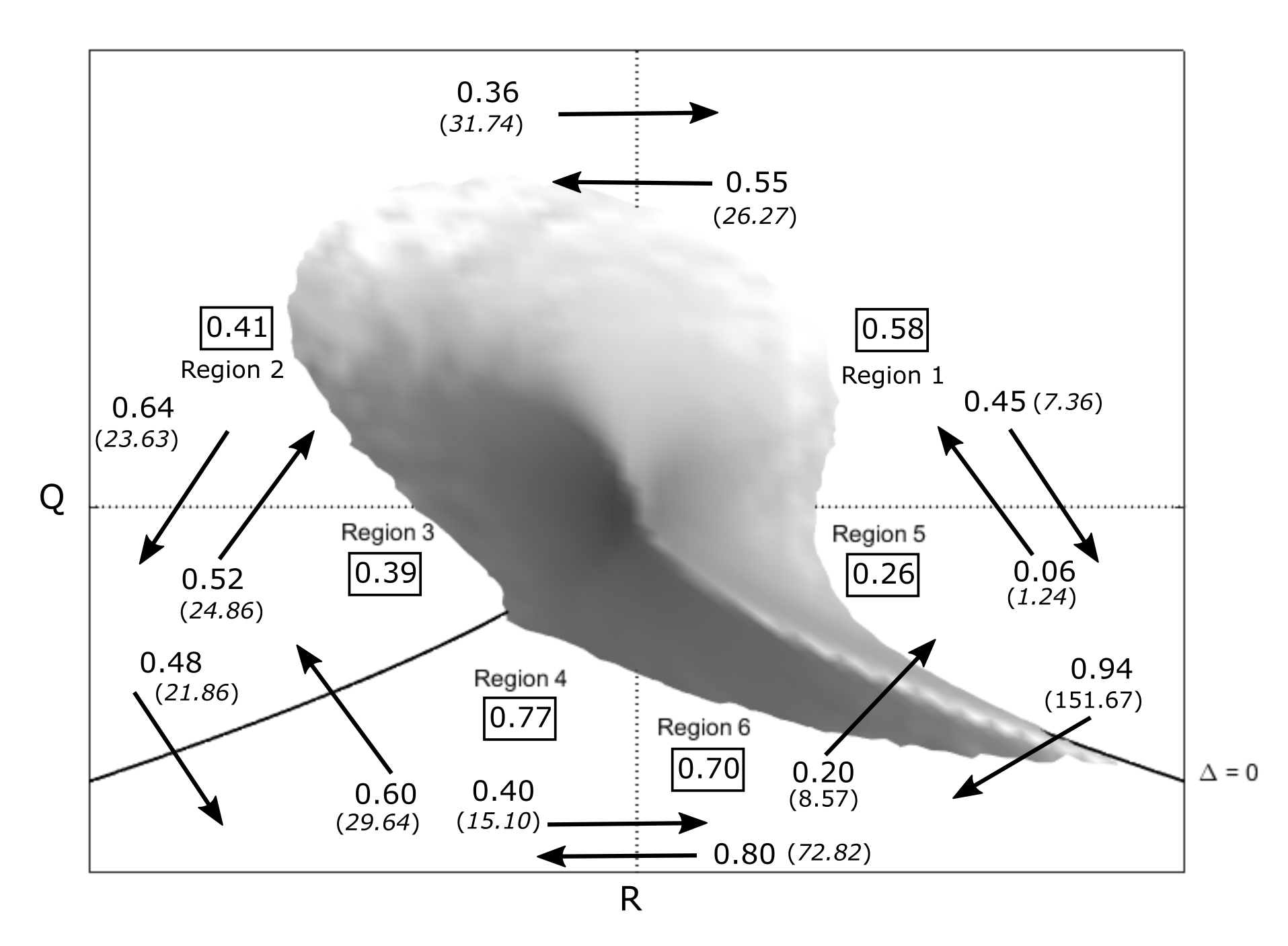}
\caption{A sketch of the joint distribution function of the VGT principal invariants, $Q$ and $R$, with the six regions and average values for the geodesics' statistics also shown. The value within a box in each region is the fraction of vertices in that region involved in inter-region geodesics. The arrows indicate the direction of the inter-region geodesics between adjacent regions, and the number next to each arrow is the proportion of the inter-region geodesics, $p_{\alpha,\gamma}^{\mathcal{S}}$, in that direction. The value in brackets is the corresponding total exit importance, $I_{\alpha,\gamma}$, defined in \eqref{eq.totimp}.}
\label{fig.QR}
\end{figure*}

Except for the long intra-region geodesics discussed above, circulation around the $Q$-$R$ diagram is primarily due to the inter-region geodesics, i.e. those where $\alpha \ne \beta$ for the path $\mathcal{S}(i,\alpha;j,\beta)$ defined in \eqref{eq.geodesic}.
Here we focus on the pair of vertices crossed at the transition from region $\alpha$ to the intermediate region $\gamma$, namely how the geodesic exits a region and enters another. Definition \eqref{eq.geodesic} gives a pair of vertices at the boundary between regions, $\{v_{k_\xi,\alpha}$ and $v_{k_{\xi+1},\gamma}\}$ for every inter-region geodesic, with $m_{k_\xi},k_{\xi+1}$ the associated weighting from $\bm{M}$.
To further characterize the fluxes between the $Q$-$R$ regions, we define the \textit{exit node importance} as 
\begin{equation}
I_{k_\xi,\alpha; k_{\xi+1},\gamma} = n_{k_\xi,\alpha}^{\mathcal{S}} m_{k_\xi,k_{\xi+1}}
\label{eq.imp}
\end{equation}
where $n_{k_\xi,\alpha}^{\mathcal{S}}$ is the number of inter-region geodesics through the region exit node $v_{k_\xi,\alpha}$ and $m_{k_\xi,k_{\xi+1}}$ is the corresponding edge weight between the boundary nodes $v_{k_\xi,\alpha}$ and $v_{k_{\xi+1},\gamma}$. 
Based on the local quantity \eqref{eq.imp} we define the total exit node importance for a region as the sum of the exit node importance over all the exit nodes from region $\alpha$ to $\gamma$,
\begin{equation}
I_{\alpha,\gamma} = \sum_{k_\xi,k_{\xi+1}} I_{k_\xi,\alpha; k_{\xi+1},\gamma}.
\label{eq.totimp}
\end{equation}

Figure \ref{fig.QR} shows a sketch of the $Q$-$R$ diagram reporting inter-region geodesic statistics. The value within a box for each region indicates the proportion of the $N_{\alpha}$ nodes in that region which are inter-region geodesic exit nodes, $v_{k_\xi},\alpha$. Within a given region $\alpha$, there are $N_{\alpha}^{\mathcal{S}} = N_{\alpha}(N-N_{\alpha})$ outgoing or incoming inter-region geodesics, as the arrows in Fig. \ref{fig.QR} show. The value at the foot of each arrow is the proportion, $p_{\alpha,\gamma}^{\mathcal{S}}$, of the inter-region geodesics that follow the direction indicated. The total exit node importance \eqref{eq.totimp} from $\alpha$ to $\gamma$ is shown in brackets next to each arrow in Fig.~\ref{fig.QR}.

The values for $p_{\alpha,\gamma}^{\mathcal{S}}$ are largely a function of $N_{\alpha}$ (reported in Table \ref{table.QR}). However, $p_{5,\gamma}^{\mathcal{S}}$ is two-thirds of $p_{3,\gamma}^{\mathcal{S}}$ implying there are additional dependencies at work. Region 5 also has the strongest asymmetry between the number of incoming and outgoing geodesics, leading to a strong directionality as  $p_{5,6}^{\mathcal{S}}\gg p_{6,5}^{\mathcal{S}}$ and $p_{1,5}^{\mathcal{S}} \gg p_{5,1}^{\mathcal{S}}$. The vertices between regions 6 and 4 also show a strong incoming/outgoing asymmetry, with $p_{6,4}^{\mathcal{S}} \gg p_{4,6}^{\mathcal{S}}$, highlighting that the regions adjoining the right Vieillefosse tail are those driving the bias towards a clockwise set of shortest paths around the $Q$-$R$ diagram. While regions 3 and 4 also exhibit such a bias towards the clockwise direction, this is not the case for regions 1 and 2 where $p_{1,2}^{\mathcal{S}}>p_{1,5}^{\mathcal{S}}$ and $p_{2,3}^{\mathcal{S}}>p_{2,1}^{\mathcal{S}}$. Furthermore, the value for $I_{\alpha,\gamma}$ exiting a region in the clockwise direction is greater than the anti-clockwise value for all regions except for region 1 where $I_{1,2} = 26.27$ and $I_{1,5} = 7.36$. Thus, there is something particularly unique about the interface between regions 1 and 2, in particular. This is in agreement with the earlier observation that the deviatoric part of the pressure Hessian generates probability fluxes that differ significantly from the clockwise restricted Euler component in region 1 \citep{chevillard08,xiao21}. By far the largest total exit importance value is $I_{5,6} = 151.7$, i.e.~across the right Vieillefosse tail. The eigenvector centrality results in Fig.~\ref{fig.centrality} and the more detailed investigation of the geodesics in Fig.~\ref{fig.QR} both draw particular attention to the vertices in the regions adjacent to the right Vieillefosse tail as being critical for the VGT dynamics and this is reinforced here.

\begin{figure*}
\includegraphics{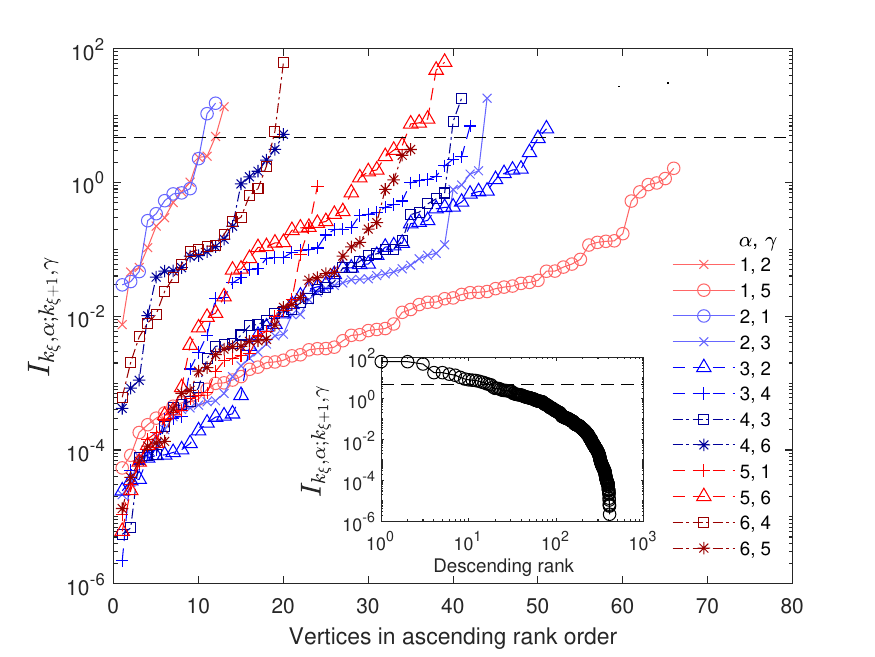}
\caption{Exit importance $I_{k_{\xi},\alpha; k_{\xi+1},\gamma}$ for each exit node of the inter-region geodesics. Each line shows the exit importance for different combinations of $Q$-$R$ regions, labeled by $\alpha$ and $\gamma$, as given in the legend, with vertices placed in ascending rank order. The sum of the values on each curve at fixed $\alpha$ and $\gamma$ gives $I_{\alpha,\gamma}$ reported in brackets in Fig.~\ref{fig.QR}. The inset shows all 407 nodes from the main panel placed in descending rank order. The nodes placed above the horizontal, dashed line in both panels account for more than 75\% of $\sum_{\alpha,\gamma}I_{\alpha,\gamma}$ over all the 407 nodes.}
\label{fig.interreg}
\end{figure*}

Moving beyond the summary values in Fig.~\ref{fig.QR}, 
we plot the exit node importance $I_{k_{\xi},\alpha; k_{\xi+1},\gamma}$ in Fig.~\ref{fig.interreg} to examine the vertices through which the transitions between $Q$-$R$ regions take place. The vertices $v_{k_\xi}$ on the horizontal axis are labelled according to their ascending rank order of importance $I_{k_{\xi},\alpha; k_{\xi+1},\gamma}$. The different lines and symbols indicate the different starting region $\alpha$ and intermediate region $\gamma$.
The inset plots all the 407 nodes $v_{k_{\xi}}$ from the main panel ordered according to their importance $I_{k_{\xi},\alpha; k_{\xi+1},\gamma}$ in descending order, using a log-scale and on a single curve. The horizontal dashed line indicates 75\% of the overall total exit importance $I_{\alpha,\gamma}$, which is given by just seventeen nodes. In the main panel of Fig. \ref{fig.interreg} we draw particular attention to the $\alpha = 1$ results, which were highlighted as unusual in Fig. \ref{fig.QR}. The mechanism by which $I_{1,2} > I_{1,5}$ is apparent from this figure: there are more vertices that serve as exit nodes in the clockwise direction from region 1 to 5 (the largest number for any region at 66) than in the anti-clockwise direction (13 nodes). However, the greatest value for $I_{k_{\xi},1; k_{\xi+1},5}$ is just 1.63 while for $\alpha = 1, \gamma = 2$ there are two instances above the threshold given by the dashed line with $I_{k_{\xi},1; k_{\xi+1},2} = 13.59$ and $I_{k_{\xi},1; k_{\xi+1},2} = 4.87$. Given that the deviatoric pressure Hessian drives the counter-clockwise trajectories, these results would suggest that this contribution to the dynamics is to open up a small number of paths that are very active, rather than the weaker transition probabilities through a large number of nodes that arises in the clockwise direction associated with the restricted Euler dynamics. 

Figure \ref{fig.interreg} also shows that there are three particularly large values for $I_{k_{\xi},\alpha; k_{\xi+1},\gamma}$ and these occur in regions 5 and 6, driving the large values in brackets for $I_{5,6}$ and $I_{6,4}$ in Fig. \ref{fig.QR}. Concerning the discrete states for these nodes, with reference to \eqref{eq.state}, the two nodes in region 5 with high exit importance are very similar: they have an intermediate ranking for $\Vert\bm{\Omega}_{C}\Vert^{2}$, positive signs of $-\mbox{det}(\bm{S})$ and $\mbox{tr}(\bm{\Omega}^{2}\bm{S}$, a positive sign for $\mbox{tr}(\bm{\Omega}_{C}^{2}\bm{S}_{B})$,
and the same ranking for the production terms.
Hence, they differ only in terms of the sign of $-\mbox{det}(\bm{S}_{C})$. The particularly important node in region 6 has $\Vert\bm{\Omega}_{C}\Vert^{2}$ as the largest in magnitude of the second-order terms with, of the production terms, the interaction production largest in magnitude, followed by the non-normal production. With $\Vert\bm{\Omega}_{C}\Vert^{2} > \Vert\bm{S}_{B}\Vert^{2}$ and $\mbox{tr}(\bm{\Omega}_{C}^{2}\bm{S}_{B}) > -\mbox{det}(\bm{S}_{C})$ it implies that 
the normal strain and the non-normality are well aligned, indicating the importance of strain and vortical alignments \citep{betchov56}, although here we are interested in the normal part of the former and the non-normal part of the latter.

\subsection{Commute distances}
The dynamics across the $Q$-$R$ plane can be characterized using geodesics, as in the previous sections or, as noted in the introduction using random walks over the network, which permits less frequented pathways to be integrated into analysis.
To this end, we studied average commute distances, i.e.~we determined the mean length $\overline{\ell}_{C}(i,j)$ of the random walks from node $i$ and back to $i$ via node $j$, for all $i$ and $j$. The averaging is undertaken over at least 20 random walk realisations for a given $(i,j)$ node pair, with a median of 3800 realisations, the variability a consequence of the orders of magnitude of variation in out degree seen in Fig. \ref{fig.OutD}b.

\begin{figure*}
\includegraphics{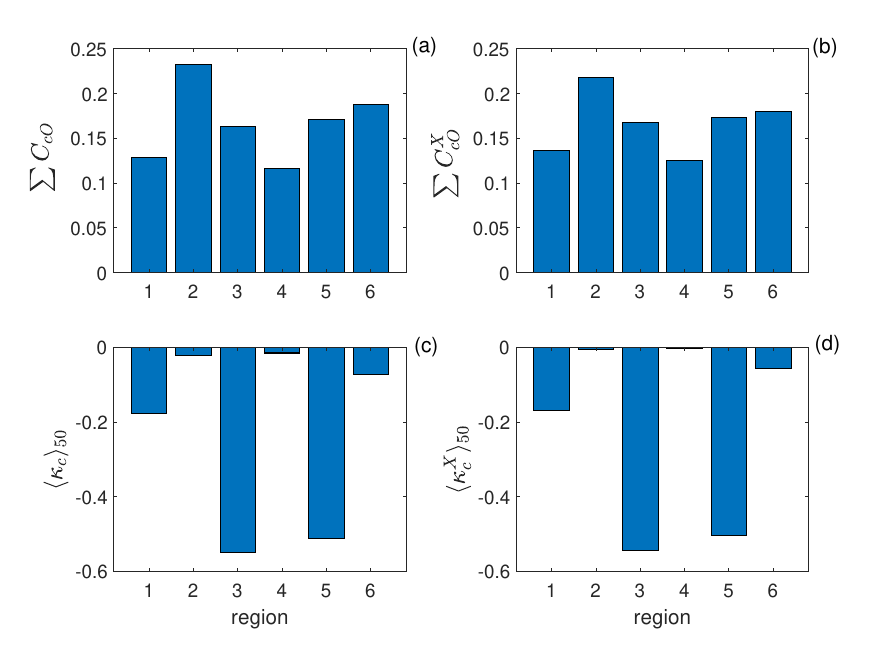}
\caption{The commute distance out-degree eigenvector centrality, $C_{cO}$, based on matrix $\bm{G}$, defined in \eqref{eq.gmat}, (a) and based on $\bm{G}^{X}$ (b). Panel (c) and (d) show the median of the asymmetry metrics $\kappa_{c}$ and $\kappa_{c}^{X}$, respectively, for each region of the $Q-R$ diagram. These results may be compared to the equivalent panels in Fig.~\ref{fig.centrality}.}
\label{fig.evecCommute}
\end{figure*}

To focus on paths that have traversed a large portion of the $Q$-$R$ diagram, we consider the intermediate node $j$ to be in the regions non-adjacent to the region $\alpha$ to which the starting/arrival node $i$ belongs.
We then define $\ell_{C}(i,j_{\min})$ as the mean commute distance for node $i$ in $\alpha$ to $i$ via an intermediate node $j_{min}$ in region $\gamma$ that gave the shortest mean commute of all the $j$ in region $\gamma$:
\begin{equation}
\ell_{C}(i,j_{\min}) = \min_j\left(\langle\overline{\ell}_{C}(i,\alpha;j,\gamma)\right), 
\label{eq.minlC}
\end{equation}
Consequently, $\ell_{C}(i,j_{\min})$ measures how well node $i$ is connected to region $\gamma$. Furthermore, in addition to characterizing the shortest mean commute, we can define the matrix $\bm{G}$ with elements
\begin{align}
\nonumber
     g_{i,j} &= 1/\overline{\ell}_{C}(i,j) \,\,\, \forall\, i \ne j\\
     g_{i,i} & = 0 \,\,\, \forall\, i = j,
\label{eq.gmat}
\end{align}
such that $\bm{G}$ has a similar interpretation to $\bm{M}$ but for commute distances rather than transition probabilities. This means we can generate \textit{commute distance eigenvector centralities} based on $G$, in an analogous way to the terms in \eqref{eq.kappa}. Namely, we can form the commute distance out-degree eigenvector centrality $C_{cO}(i)$, the in-degree centrality $C_{cI}(i)$ and their normalized difference $\kappa_{c}(i)$. Furthermore, in the same way that $\bm{M}^{X}$ in (\ref{eq.mXdef}) is a special case of $\bm{M}$, restricting attention to edges between nodes in different regions, we may also form $\bm{G}^{X}$ in an analogous fashion and undertake an analysis that directly parallels that shown in Fig. \ref{fig.centrality} but for the commute distance eigenvector centralities.

We show the values of $C_{cO}$ and $\kappa_{c}$ for each $Q$-$R$ region in Fig.~\ref{fig.evecCommute}, which is directly analogous to Fig.~\ref{fig.centrality}. The most noticeable difference is that in Fig.~\ref{fig.centrality} the results radically changed based on whether $\bm{M}$ or $\bm{M}^{X}$ underpinned the analysis, while here, using $\bm{G}$ or $\bm{G}^{X}$ results in only mild changes. This is primarily for two reasons: random walks are innately more dispersive than geodesics meaning that the probability of venturing out of a region for an intra-region commute is higher; and, because the commute distance requires a return to a node, the percentages quoted in the left-hand panels in Fig.~\ref{fig.ellSN} would innately be greater even if we were considering a returning geodesic rather than a random walk. In comparison to the results for $C_{eO}$ and $C_{eO}^{X}$ where regions 5 and 6 dominated the statistics, here we see that region 2 has the greatest value for $C_{cO}$. However, the strongly negative values for $\langle\kappa_{c}\rangle_{50}$ in regions 3 and 5 mean that these regions have large $C_{eI}$ and in contrast to Fig. \ref{fig.centrality}, no regions have, on average, $C_{eO}^{X}$ exceeding $C_{eI}^{X}$. Hence, in-degree clustering of the vertices based on short commute distances is more prevalent than for the conventional eigenvector centrality.    

\begin{figure*}
\includegraphics{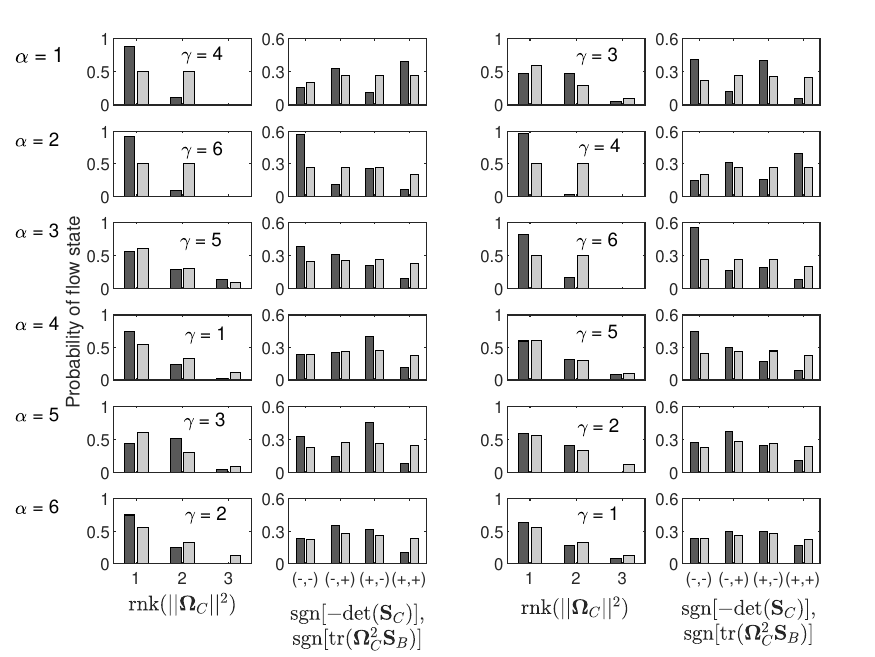}
\caption{The probability of different flow states for nodes, $j_{\min}$ that form the shortest mean commutes, $\ell_{C}(i,j_{\min})$ in region $\gamma$ (dark gray) compared to all the vertices in a region (light gray). For a region $\alpha$ containing the node $i$ we show results for two regions $\gamma$ as indicated in the left-hand panel of each pair, which are three and four regions apart from the starting region $\alpha$ based on a clockwise motion around the $Q$-$R$ diagram. For each pair of panels, the left-hand panel shows the ranking of the non-normality, $\Vert\bm{\Omega}_{C}\Vert^{2}$, while the right-hand panel shows the signs of the non-normal production and interaction production. 
}
\label{fig.OmC2}
\end{figure*}

Given that the $\ell_{C}(i,j_{\min})$ provide the shortest commute distances for a node $i \in \alpha$ to any of the nodes in region $\gamma$, we examine the characteristics of the $j_{\min}$ nodes (dark gray bars) relative to all nodes in the same region (light gray bars) in Fig. \ref{fig.OmC2}. The regions of concern are indicated by the value for $\alpha$ for node $i$ (each row of panels) and then the value for $\gamma$ for the nodes $j$ (or $j_{\min}$ stated in the left-hand panel of each pair. We focus on three aspects of the flow state vector for each vertex: the relative ranking of the magnitude of $\Vert\bm{\Omega}_{C}\Vert^{2}$ (compared to $\Vert\bm{\Omega}_{B}\Vert^{2}$ and $\Vert\bm{S}_{B}\Vert^{2}$) in the left-hand panels, and the signs of the non-normal production and the interaction production in the right-hand panels of each pair. These are the same terms considered earlier to explain the bimodality of intra-region geodesics for regions 1 and 5. We focus on two regions $\gamma$ for each $\alpha$, which are those that are three or four regions apart from $\alpha$ assuming a clockwise circulation around the $Q$-$R$ diagram. 
For commutes from the $Q > 0$ regions ($\alpha \in \{1,2\}$) via vertices within $\Delta < 0$ regions ($\gamma \in \{4,6\}$) or \textit{vice versa}, and particularly for $\alpha = 2$ and $\gamma \in \{4,6\}$, there is a clear preference for $\Vert\bm{\Omega}_{C}\Vert^{2}$ to be the largest term in magnitude at the $j_{\min}$ node. Based on a chi-squared test, all of the differences in $\Vert\bm{\Omega}_{C}\Vert^{2}$ between the $j_{\min}$ nodes and all the nodes in $\gamma$ were highly significant for all regions apart from the $\alpha = 6, \gamma = 1$ case. Thus, non-normality plays an important role in facilitating short commutes around $Q$-$R$ space despite not appearing in either $Q$ or $R$, implying the need to consider additional terms to those in the restricted Euler model to understand the Lagrangian VGT dynamics more completely \citep{chevillard08,luthi09,Johnson2020,Tom2021}.

The two regions where $Q < 0$ but $\Delta > 0$, i.e.~regions 3 and 5, all exhibit significant differences concerning the signs of the non-normal production and the interaction production whether they are a source node, $\alpha$, or an intermediate node, $\gamma$. In particular, where $\gamma = 3$, which is the case for $\alpha \in \{1,5\}$, such results are driven by $\mbox{tr}(\bm{\Omega}_{C}^{2}\bm{S}_{B}) < 0$, as the $j_{\min}$ cases have an excess of either both terms negative or just the interaction production. In contrast, where $\gamma = 5$, which is the case for $\alpha \in \{3,4\}$, the $j_{\min}$ cases are driven by $-\mbox{det}(\bm{S}_{C}) < 0$. 
It was already shown in Table \ref{table.reg5} that intra-region geodesics which traverse the $Q$-$R$ space are associated with negative values for non-normal production and, in particular, interaction production. The fact that the short commute distances are also aided by negative non-normal and interaction production further highlights the importance of these flow characteristics. We note that the flow states that facilitate rapid commute times are relatively unusual since positive values of $\mbox{tr}(\bm{\Omega}_{C}^{2}\bm{S}_{B})$ generally dominate on average as seen in the last column of Table \ref{table.reg5} and in \cite{k18}.

\section{Discussion}
\begin{figure}
\includegraphics{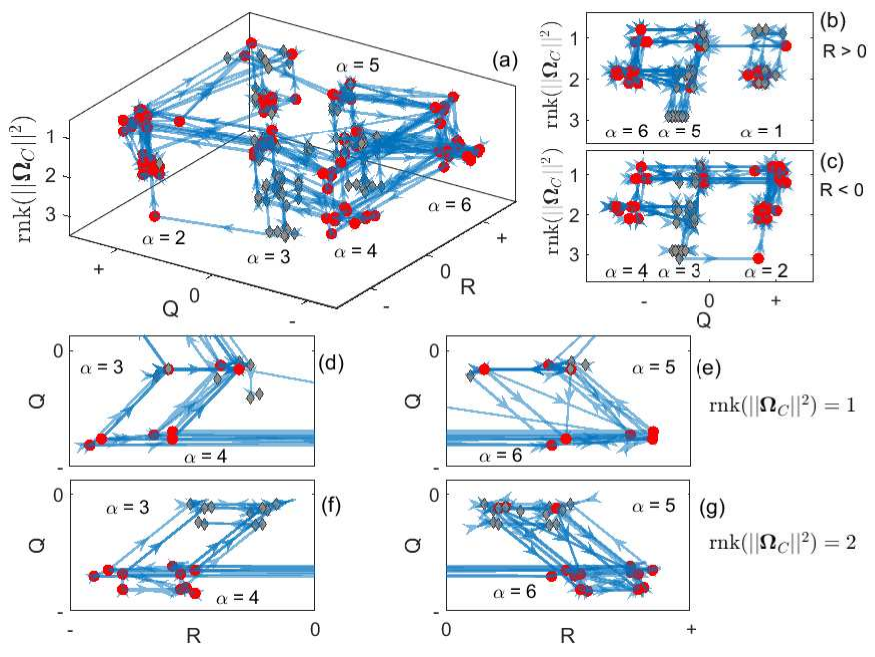}
\caption{A visualization of the sub-network constructed from $\bm{M}$ using the 178 nodes with highest out-degree is shown in full in panel (a). Each node is placed in its appropriate $Q$-$R$ region, with the vertical axis indicating the ranking of non-normality, $\Vert\bm{\Omega}_{C}\Vert^{2}$, compared to the enstrophy and total straining (see definition \eqref{eq.state}). The red circle nodes are those in the top 10\% based on their out-degree, while the gray diamonds are the others that meet the additional selection criterion (see text). We show the edges with a weight exceeding a given threshold, and with the arrowhead indicating the net direction of the transition probability. Panels (b) to (g) emphasize sub-components of the main panel. Nodes belonging to the same region are displaced from one another horizontally and vertically to aid visualization.}
\label{fig.netw}
\end{figure}

Aspects of the network structure examined above can be appreciated in the sub-network of $\bm{M}$ consisting of 178 nodes shown in Fig.~\ref{fig.netw}. These nodes were selected based on their out-degree: the largest 10\% (84 nodes) were initially selected (shown as red circles); we then extracted the next largest nodes until the total number of nodes selected for a region equalled or first exceeded $0.2 N_{\alpha}$. The 178 nodes extracted (21\% of the total) accounted for 56\% of the total out-degree over all nodes. Edges are included in Fig.~\ref{fig.netw} where 
$m_{i,j} > 75 / N(N-1)$ 
and the line thickness is proportional to the logarithm of $m_{ij}$. Further simplification for visualization purposes is achieved by only showing a single edge between vertices; if both $m_{ij}$ and $m_{ji}$ exceed the threshold, only the larger is shown as indicated by the direction of the arrow. Because $\Vert\bm{\Omega}_{B}\Vert^{2} = 0$ in regions 4 and 6 there are no nodes at $\mbox{rnk}(\Vert\bm{\Omega}_{C}\Vert^{2}) = 3$ by definition. No such constraint arises for region 1, but panel (b) highlights that there are no nodes in this region with sufficiently large out-degrees to be shown. Note this relates directly to our earlier result about the bimodality of intra-region geodesics and the need in region 1 for nodes to gain $\Vert\bm{\Omega}_{C}\Vert^{2}$ to connect to major pathways. The VGT cannot gain non-normality $\Vert\bm{\Omega}_{C}\Vert^{2}$ within region 1, resulting in a circuit around the $Q-R$ diagram taking less well-frequented pathways that are not shown in Fig. \ref{fig.netw}. In addition, relatively few edges connect region 1 to region 5, reflecting the results in Fig.~\ref{fig.interreg} where the stronger pathways were counter-clockwise (from region 1 to 2).

Figure ~\ref{fig.interreg} shows that one node in region 2 ($Q>0$, $R<0$) has both a minimal ranking for the non-normality, $\mbox{rnk}(\Vert\bm{\Omega}_{C}\Vert^{2}) = 3$, and large out-degree (within the largest 10\%). In contrast, none of the nodes with $\mbox{rnk}(\Vert\bm{\Omega}_{C}\Vert^{2}) = 3$ in regions 3 or region 5 have very large large out-degree (not within the largest 10\%, they are all shown as gray diamonds). Therefore, this node in region 2 with small non-normality and large out degree plays a unique role in the highlighted network as it provides a path to move clockwise from region 3 to region 2 with low non-normality. 
Given that $\Vert\bm{\Omega}_{B}\Vert^{2} = 0$ in regions 4 and 6, as the VGT moves clockwise and crosses the left Vieillefosse tail its eigenvalues become complex and the VGT gains some $\Vert\bm{\Omega}_{B}\Vert^{2}$ going from regions 4 to region 3. In region 3, the normal enstrophy production is positive since $R < 0$, and $\Vert\bm{\Omega}_{B}\Vert^{2}$ develops rapidly such that the ranking of the non-normality changes from $\mbox{rnk}(\Vert\bm{\Omega}_{C}\Vert^{2}) = 2$ to $\mbox{rnk}(\Vert\bm{\Omega}_{C}\Vert^{2}) = 3$. The node in region 2 with low non-normality we have discussed then provides a means for tensors with low non-normality to progress clockwise on the $Q$-$R$ diagram from region 3 to region 2. These are coherent flow structures in the sense of \citep{xu19} as they have a strong solid-body rotation and low non-normality. When the flow arrives in region 2, the straining declines and the non-normality increases to $\mbox{rnk}(\Vert\bm{\Omega}_{C}\Vert^{2}) = 2$. Panel (c) shows that the other major pathways from region 3 to 2 are all where $\mbox{rnk}(\Vert\bm{\Omega}_{C}\Vert^{2}) = 1$, so this coherent flow structure pathway is rather novel as other pathways require strong non-normality, a feature of the VGT dynamics outside the $Q$-$R$ plane \citep{luthi09}.

The densest connections are between regions 5 and 6 and a comparison of panels (e) and (g) in Fig.~\ref{fig.netw} shows that the strongest connections arise for $\mbox{rnk}(\Vert\bm{\Omega}_{C}\Vert^{2}) = 2$ rather than $\mbox{rnk}(\Vert\bm{\Omega}_{C}\Vert^{2}) = 1$. Panels (e) and (g) include edges pointing in a counter-clockwise direction. This indicates that the VGT state can oscillate between these two $Q$-$R$ regions, driving the large eigenvector centralities seen in Table \ref{table.QR} and the well-known concentration of the VGT state along the Vieillefosse tail \citep{Cantwell1993,vieillefosse84,laizet15}. 
Finally, panels (a-c) in Fig.~\ref{fig.netw} indicate that the major transfers between regions where $Q < 0$ take place at a constant ranking of the non-normality. Hence, regions 3 and 5 are crucial in providing ``vertical'' pathways for re-organizing the importance of the non-normality relative to the normal straining, which will be greater than the normal enstrophy in these regions. This explains why these two regions have the smallest values of the median centrality $\langle C_{b}^{X}\rangle_{50}$ in Table \ref{table.QR} as they are focused on internal flow reconfiguration rather than transitions to other regions. In region 5, there is no entrance or exit pathway of sufficient magnitude to be displayed for $\mbox{rnk}(\Vert\bm{\Omega}_{C}\Vert^{2}) = 3$, resulting in a significant intra-region dynamics with only a small proportion of nodes involved in inter-region geodesics (cf.~the values reported in boxes in Fig.~\ref{fig.QR}). 

\section{Conclusion}
Representing the Lagrangian dynamics of the velocity gradient tensor as a complex network has allowed us to explore the VGT time evolution conditional on several discrete variables, something challenging to achieve in a statistically robust manner in a continuous framework.
The complex network associated with the VGT dynamics comprises $N = 844$ nodes, each representing a different flow state according to the sign and relative ranking of relevant VGT invariants. Those invariants follow from a complex Schur decomposition of the VGT and consist of combinations of enstrophy and strain-rate magnitude, together with the associated production terms, accounting for the contributions from the normal and non-normal parts of the tensor (see definition \ref{eq.state}). Edges in the network capture the probability of transitioning between the discrete flow states, following the VGT Lagrangian evolution along fluid particle trajectories in a direct numerical simulation of statistically isotropic turbulence \citep{yili}.

We have analysed the network using a variety of well-established metrics of network structure \citep{newman03,Latora2017} as well as variants specifically defined to focus on the transitions between nodes belonging to different regions of the $Q$-$R$ plane. We complemented this perspective by considering the geodesics on the network, conditional on their starting and arrival region, discerning between those starting and ending in the same $Q$-$R$ region (intra-region geodesics) and those originating and ending in distinct $Q$-$R$ regions (inter-region geodesics).
The eigenvector centrality shows preferential clustering of the nodes visited with high probability on the two regions adjacent to the right Vieillefosse tail \citep{Cantwell1993,chacin00}. 
This is particularly the case for region 6, immediately below the Vieillefosse tail (where $Q < 0$, $R > 0$, $\Delta < 0$) as seen in Fig. \ref{fig.centrality}a. This indicates a strong clustering of the nodes with high connectivity. Such a strong bias towards this region is reduced when the network is constructed in terms of edges between regions only (Fig. \ref{fig.centrality}b), indicating the high eigenvector centrality for the full network is driven by connections within this region and an associated reorganization of flow characteristics without a change in region.

The remaining two regions with $R > 0$ (regions 1 and 5) were particularly unusual in terms of having particularly long paths for some of their intra-region geodesics. The reason for this was that there were certain flow characteristics that could not be readily gained without a change in region. Instead, it was necessary to undertake a full orbit of the $Q$-$R$ diagram following relative low probability pathways to gain these flow properties, resulting in surprisingly long geodesics. The driving mechanism for this differed for the two regions. In region 1 ($Q > 0$, $R > 0$) these trajectories arose for flow states with low non-normality, $\Vert\bm{\Omega}_{C}\Vert^{2}$, which needed this term to increase significantly in order to access the high probability path to region 5 (see Fig. \ref{fig.netw}b). In region 5, the changes required were associated with the production of the non-normality by the action of the normal straining, with the long, low probability paths associated with gaining negative values for this production term. 
These long geodesics are associated with a marked bimodality of the geodesics' length distribution in regions 1 and 5 and these are the $Q$-$R$ regions in which the deviatoric pressure Hessian introduces convoluted probability fluxes, very different from the underlying restricted Euler part \citep{cantwell92,chevillard08,zhou15}. Furthermore, the reconfiguration of the non-normal parts of the VGT following the long intra-region geodesics highlight the role of additional dynamical terms at play, the deviatoric pressure Hessian and viscous contributions, that we cannot capture by considering solely the restricted Euler dynamics and the resulting $Q$-$R$ trajectories.

Additional emphasis on the non-normality comes from the random walk commute distances from node $i$ and back to $i$ via node $j$. The nodes $j_{\min}$, responsible for the shortest commutes preferentially feature the non-local rotation rate $\bm{\Omega}_{C}$ larger in magnitude than either the normal straining, $\bm{S}_{B}$, or the normal rotation rate, $\bm{\Omega}_{B}$. Such vertices with large non-normality form the upper layer in the simplified network shown in Fig.~\ref{fig.netw}, and the edges between such nodes are of particular importance for the transitions from the $Q < 0$ to $Q > 0$ regions, as highlighted in panels (b) and (c) of that figure. The non-normality $\Vert\bm{\Omega}_{C}\Vert^{2}$ does not feature explicitly in the expression of the VGT principal invariants as seen from equations \eqref{eq.Om} and \eqref{eq.trOmS}, demonstrating the importance of extending the discrete and qualitative description of the VGT outside the $Q$-$R$ diagram. A previous approach in this direction due to \citep{luthi09} considered $Q$ together with the strain production and enstrophy production separately. The Schur decomposition of the VGT can enrich the decomposition by including the separate effects of $\mbox{tr}(\bm{\Omega}_{B}^{2}\bm{S}_{B})$, $-\mbox{det}(\bm{S}_{B})$, $-\mbox{det}(\bm{S}_{C})$ and $\mbox{tr}(\bm{\Omega}_{C}^{2}\bm{S}_{B})$. The signs of the additional invariants are key for explaining the long intra-region geodesics starting from region 5, together with the resulting bimodal distribution of the length of the geodesics. However, they cannot explain the surprisingly long intra-region geodesics and random walk commutes starting from region 1, for which the crucial quantity is the non-normality $\Vert\bm{\Omega}_{C}\Vert^{2}$.

Translating an infinite number of continuous flow states into a network of a finite number of discrete states provides several advantages to tackle the multi-scale complexity of turbulence \citep{wilczek24}. It constitutes a viable approach for elucidating the physics of the VGT through the application of network theory, also opening up the potential to use discrete control formulations for, e.g., turbulent drag reduction. Moreover, the complex network analysis highlighted the crucial role of the non-normal terms such as $\Vert\bm{\Omega}_{C}\Vert^{2}$ which are strongly affected by the deviatoric part of the pressure Hessian and thus related to the non-locality of the flow \citep{ohkitani95}. The edges linking the VGT discrete states in the complex network encode, visually and intuitively, the complex action of the non-local terms on the VGT dynamics.

\acknowledgments{The authors acknowledge Michael Wilczek for fruitful discussions about this work. 
CK acknowledges financial support from Leverhulme Trust International Fellowship 2023-014 that permitted the visit to Bayreuth that enabled this research.}

%

\end{document}